\documentclass[doublespacing]{bmcart}

\usepackage[utf8]{inputenc} 
\usepackage{mathtools, amsmath, amssymb, graphicx, amsfonts}
\usepackage{xcolor, comment} 
\usepackage{bbm} 
\usepackage{float} 
\usepackage{amsopn} 
\usepackage{lastpage,fancyhdr,graphicx}

\begin{document}

\begin{frontmatter}

\begin{fmbox}
\dochead{Research}

\title{Stroke recovery phenotyping through network trajectory approaches and graph neural networks}

\author[
  addressref={aff1},                   
  corref={aff1},                       
  email={s.krishnagopal@ucl.ac.uk}   
]{\inits{S.K.}\fnm{Sanjukta} \snm{Krishnagopal}}
\author[
  addressref={aff2},                   
  email={lohse@wustl.edu}   
]{\inits{K.L.}\fnm{Keith} \snm{Lohse}}
\author[
  addressref={aff3,aff4},                   
  email={Robynne.Braun@umm.edu}   
]{\inits{R.B.}\fnm{Robynne} \snm{Braun}} 


\address[id=aff1]{
\orgdiv{Gatsby Computational Neuroscience Unit},
  \orgname{University College London}, 
  \postcode{W1T 4JG},                              
  \city{London},                              
  \cny{UK}                                    
}

\address[id=aff2]{%
  \orgdiv{Physical Therapy and Neurology},
  \orgname{Washington University School of Medicine},
  \street{4444 Forest Park Ave., Suite 1101},
  \postcode{63108-2212}
  \city{St. Louis, MO},
  \cny{USA}
}

\address[id=aff3]{%
  \orgdiv{Department of Neurology, Baltimore, Bressler Research Building},
  \orgname{University of Maryland},
  \street{655 W. Baltimore Street, 12th Floor},
  \postcode{21201}
  \city{Baltimore, MD},
  \cny{USA}
}

\address[id=aff4]{%
  \orgdiv{GPAS Collaboration, Phenotyping Core},
}

\end{fmbox}

\begin{abstractbox}

\begin{abstract} 
	
	Stroke is a leading cause of neurological injury characterized by impairments in multiple neurological domains including cognition, language, sensory and motor functions. Clinical recovery in these domains is tracked using a wide range of measures that may be continuous, ordinal, interval or categorical in nature, which presents challenges for standard multivariate regression approaches. This has hindered stroke researchers’ ability to achieve an integrated picture of the complex time-evolving interactions amongst symptoms. Here we use tools from network science and machine learning that are particularly well-suited to extracting underlying patterns in such data, and may assist in prediction of recovery patterns. To demonstrate the utility of this approach, we analyzed data from the NINDS tPA trial using the Trajectory Profile Clustering (TPC) method to identify distinct stroke recovery patterns for 11 different neurological domains at 5 discrete time points. Our analysis identified 3 distinct stroke trajectory profiles that align with clinically relevant stroke syndromes, characterized both by distinct clusters of symptoms, as well as differing degrees of symptom severity. We then validated our approach using graph neural networks to determine how well our model performed predictively for stratifying patients into these trajectory profiles at early vs. later time points post-stroke. We demonstrate that trajectory profile clustering is an effective method for identifying clinically relevant recovery subtypes in multidimensional longitudinal datasets, and for early prediction of symptom progression subtypes in individual patients. This paper is the first work introducing network trajectory approaches for stroke recovery phenotyping, and is aimed at enhancing the translation of such novel computational approaches for practical clinical application.
	
	
\end{abstract}


\end{abstractbox}
%

\end{frontmatter}


\section{Introduction}

\textbf{Stroke Recovery is Dynamic and Involves Multiple Neurological Domains.}

The process of neurological recovery after brain injuries such as stroke entails complex interactions among multiple variables that change dynamically over time \cite{wang2020efficiency, lohse2016quantifying}. It is well known that the degree of recovery after stroke varies widely between individuals \cite{STI2012,WAH2014,SIM2020,braun2021domain}, where each patient’s recovery pattern uniquely reflects the combined influence of their lesion size and location \cite{shelton2001effect}, baseline health status, time to initial treatment \cite{dromerick2015critical}, and response to medical treatment or rehabilitation, among many other intrinsic and extrinsic factors.  Recovery trajectories furthermore vary depending on the specific neurological domain(s) affected (i.e. for motor, language, or sensory impairments),\cite{CRA2007,CRA2021} and each of these symptoms may show varying responsiveness to treatment. For example, language problems (aphasia), right-sided motor symptoms, and spatial perceptual problems (hemineglect) are reportedly less responsive than other symptoms to  treatment with tissue plasminogen activator (tPA)\cite{FEL2002}.  Stroke recovery is therefore notoriously heterogeneous in terms of the type and severity of residual symptoms, as well as the timecourse of progression and/or resolution of those symptoms \cite{van2020predicting}.

An important goal for stroke research is to reduce the `noise' arising from this inherent heterogeneity by stratifying patients who are likely to have similar symptom trajectories.
The heterogeneity of symptoms and time-varying recovery patterns inherent to stroke make it an area especially well-suited area for data-driven approaches.  The increasing availability of large scale stroke datasets has led to a recent explosion in the use of data-science methods for stroke research \cite{wang2020systematic}. 
For example, machine learning analyses of stroke clinical data \cite{stinear2017prep2, van2020predicting} have been used to characterize symptom clusters\cite{SUC2013}, predict outcomes \cite{STI2012}, and define composite measures of recovery \cite{HOM2016}.

\textbf{Limitations of Conventional Regression and Machine Learning Approaches versus Network Science}

While Machine Learning (ML) approaches have been successful in a variety of analytical tasks, they often present challenges for interpretation and subsequent application. By contrast, network science tools are explicit in their modeling, making them more useful for studying medical data where clinical interpretation is paramount. Additionally, typical ML approaches focus on prediction, while taking the outcome itself at face value; in contrast, network approaches attempt to improve how the outcome itself is captured. While ML tools may not be as readily interpretable as network approaches for various types of analyses (such as understanding the interactions that underlie disease recovery patterns) ML still presents several desirable properties, particularly in terms of data-driven predictive ability, and may therefore be useful for prognostiction of patient recovery patterns. 

Conventionally, statistical tools such as mixed-effects regression are used for modeling longitudinal data in disciplines where repeated measures designs are particularly relevant, such as education, motor learning, and psychology (\cite{singer2003applied, garcia2017statistical, lohse2020modeling}). Mixed-effects models have tremendous flexibility in their ability to accommodate different types of study designs, especially for handling repeated measures and the clustering of observations within groups. Such models are thus increasingly used in fields like neurorehabilitation where serial measures of recovery constitute a central focus \cite{garcia2017statistical, kozlowski2013using, lohse2016quantifying}. Sophisticated though mixed-effects models may be as statistical tools, they face several challenges when applied to problems that commmonly arise in recovery research, such as multivariate outcomes, ordinal outcomes, and violations of modeling assumptions. Many of these challenges can be readily addressed by using network approaches to data analysis.

\textbf{Network Science and Trajectory Profile Clustering for Stroke Research} 

Insights into the complex patterns of symptom evolution can be gained through the computational power of networks analysis. The field of network medicine \cite{PAW08} studies disease manifestation and progression as a function of multiple interacting disease variables, which may be of similar or different types. Networks approaches also produce intuitive data visualizations that can facilitate interaction between clinicians and data scientists to yield novel insights on disease. However, with few exceptions, most network medicine studies have focused on biomolecular data \cite{HUA04,BAR11} rather than characterizing patients' patterns of symptom progression over time. 

Recently, Krishnagopal et al. \cite{KRI20} introduced a network-based approach called Trajectory Profile Clustering (TPC) that groups patients based on similar patterns of symptom evolution. The intuitiveness and ability of TPC to integrate variables on multiple different scales make it especially useful for studying disease severity, progression, and recovery. Multi-layer \cite{KRI20_} types of trajectory clustering have also shown success in clinically-validated disease trajectory prediction in Parkinson's disease. We argue that TPC offers unique advantages for stroke recovery research based on its ability to simultaneously (i) identify the dominant variables that differentiate stroke recovery subtypes, (ii) account for temporal disease progression patterns, and (iii) delineate distinct symptom groupings. This paper is the first work introducing network trajectory approached for stroke recovery phenotyping, and is aimed at enhancing the translation of such novel computational approaches for practical clinical application.


When analyzing recovery trajectories with TPC, an obvious question that arises is at what stage do patients begin to stratify into distinct trajectory clusters (i.e. when do they begin to show symptom patterns unique to their recovery subtype)? The timing of medical treatments might be one important influence on the timecourse of recovery subtype stratification. For example in stroke, stratification might be expected to occur based on when patients receive treatments such as tPA or clot retrieval. Naively, this may appear to be a problem of simply measuring the differences between trajectory subtypes at each timepoint.  However, since treatment efficacy for the same individual at different timepoints is not unrelated, more sophisticated tools are required to extract the timescale of separation. We can investigate these questions through graphical tools such as Graph Neural Networks in machine learning.


\textbf{Graph Neural Networks for the Study of Neurological Disorders.} 

The field of machine learning has been revolutionized by recent advances in deep neural networks, especially convolutional neural networks (CNNs) \cite{ALB17}. Conventionally, CNNs use local connections, shared weights and multiple layers to extract representations of data. However, CNNs work in a Euclidian domain and are best suited for use with images. By contrast, other deep learning methods can operate on a graph domain (i.e. graph neural networks or GNNs) \cite{SCA08}. Convolutional variants of graph neural networks provide a framework for transferring deep learning operators into a non-Euclidean (graphical) domain, and have been successful in a variety of tasks such as graph classification, node identification, link prediction in protein interactions, knowledge graphs, and social network analysis, among others. Of particular interest here, they have been successfully applied to study neurological disorders including Alzheimer's disease and autism \cite{SON19,PAR18}, and could similarly have utility in the study of stroke.

\section{Stroke Dataset Analyzed}
To demonstrate the utility of using TPC and CNN for stroke recovery research, we analyzed cases from the well-characterized NINDS tPA trial data set \cite{NIH1995}.  This study was a randomized, double-blinded, placebo-controlled trial that compared the effects of intravenous tPA (a thrombolytic agent used in ischemic stroke to dissolve blood clots) versus placebo treatment in 624 patients. The data set captures neurologic deficits on the NIH Stroke Scale (NIHSS) \cite{BRO1989}, which is the most widely used measurement scale for stroke neurologic deficits, and has clinimetric properties that are well-defined \cite{MIL2007,goldstein1997reliability}. Each item is scored on a scale (from 0-3 or 0-5), with higher values indicating greater stroke severity.  The NINDS tPA data captures NIHSS scores across 5 time points: at hospital admission, at 2 hours, 24 hours, 7-10 days, and 3 months post-stroke. Here, we examined symptom progression in 11 neurologic domains as assessed by 15 individual item subscores on the NIHSS. We excluded a total of 135 cases who had imputed data at any time point (134) and/or had died (118). We excluded these cases because the imputation approach that had been used could distort patterns of change in scores for individual patients (i.e. missing values were imputed as the worst score for each NIHSS item). After exclusions, there were 489 remaining cases for analysis.

\begin{table}[ht]
\centering 
\caption{NINDS tPA Trial Data, Variable Names and Symptom Descriptions} 
\begin{tabular}{l l} 
\hline\hline 
Variable Name & Symptom Description \\ 
\hline 
ATAXIA & Coordination \\
CONSCIO & Level of consciousness \\
DYSAR & Speech (slurring)\\
EXTIN & Spatial perception\\
GAZE & Eye movements\\
LANG & Language\\
LOCCOM & Command following\\
LOCQU & Question answering\\
MOTORLA & Left arm strength\\
MOTORLL & Left leg strength\\
MOTORRA & Right arm strength\\
MOTORRL & Right leg strength\\
PALSY & Facial weakness\\
SENSORY & Skin sensation\\
VISUAL & Visual fields\\

\hline 
\end{tabular}
\label{table:data} 
\end{table}

\begin{figure}[htbp!]
  \includegraphics[width=1\textwidth]{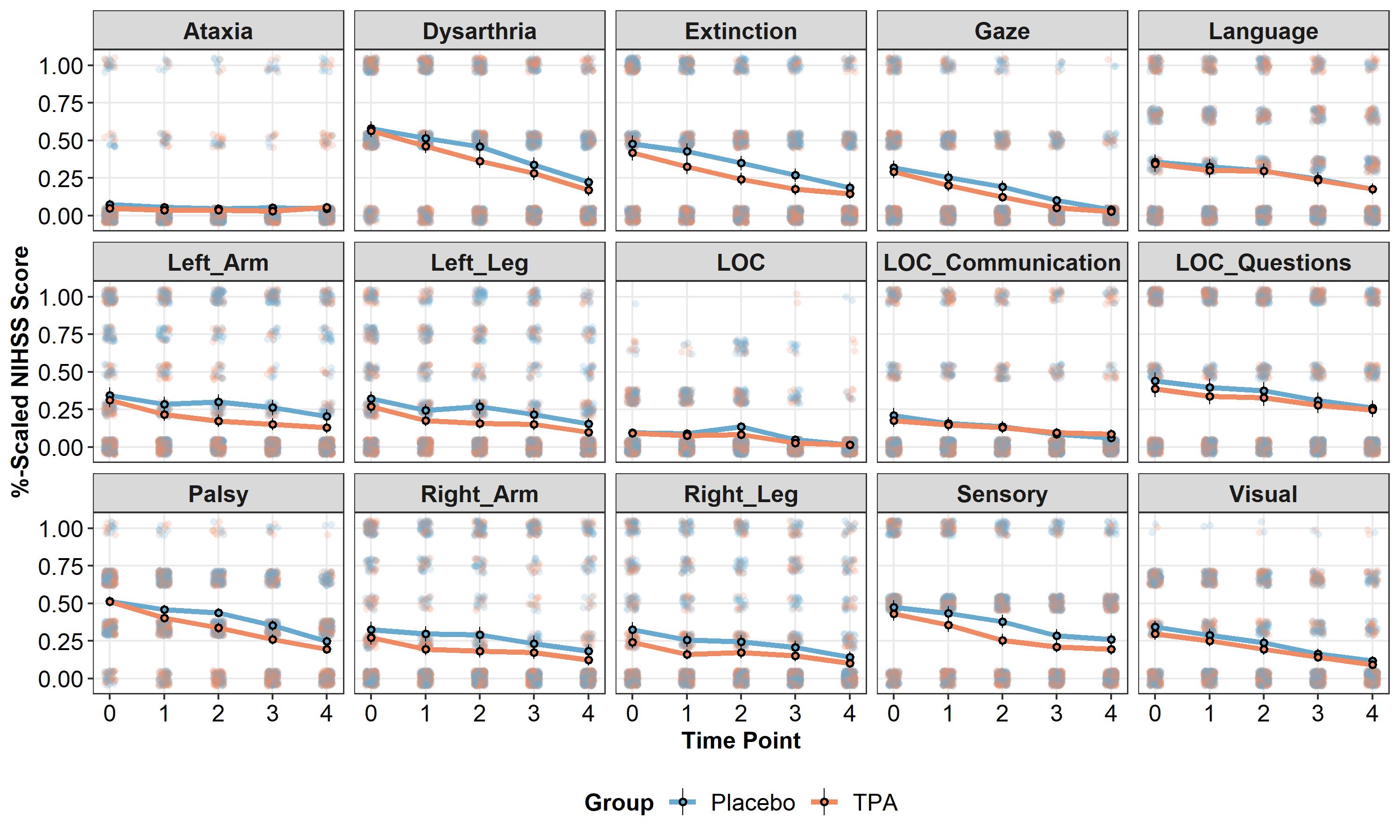}
    \caption{Data from the NINDS tPA trial shown as a function of treatment group, time, and the 15 NIHSS assessment items. Data have been scaled to be a percentage of the maximum score (on the y-axis) in each domain and have been jittered to show overlapping points. Solid lines and outlined circles show the mean performance at each time point. Error bars show the 95\% confidence intervals for each point.
    }
  \label{fig:tPA_data}
\end{figure}

\section{Methods}
\subsection{Mixed-Effects Regression Model}

The NINDS dataset contains repeated measures at 5 time points across 11 different neurological domains (as measured by 15 different NIHSS assessment items), resulting in 75 observations per patient. 
To understand how changes in different symptoms related to each other over time, we extracted the random-effects from the model to compute slopes and intercepts for each individual overall, and for each individual in each domain. Using these coefficients we could test two different hypotheses to better understand behavioral phenotypes: (1) how slopes related to intercepts both overall and within specific domains and (2) how slopes (rates of change) related to each other across domains (i.e. which symptoms improve together). Given that the NIHSS items have different maxima (all have a minimum of 0), we choose to standardize NIHSS scores to be a fixed percentage of the maximum in each domain. Fig. \ref{fig:tPA_data} shows these standardized data for each NIHSS assessment item as a function of time. Inspection of the raw data revealed substantial right skew in this dependent variable, so we log-transformed the standardized values ($log_{10}(x_{hij}/max(x_{h})+1)$), to achieve a more normal distribution of scores. (Note, $h$ indexes the scale, $i$ the individual participant, and $j$ the specific time-points). 

To account for the statistical dependence among these repeated measures, we include random-intercepts and random-slopes for the effect of `Time' at both the level of patient, and at the level of patient:domain. The technical details are outlined here. Conceptually, effects at the patient level reflect the average change for a patient over time across domains, represented by a unique slope and intercept for each individual, $(1+Time|patient)$. Effects at the domain-level reflect the change over time in each domain for a particular patient, which can be captured by a unique slope and intercept in each domain for each individual, $(1+Time|(patient:domain))$. Models also includes fixed-effects of Time, Group (tPA versus placebo), Domain (the 15 assessment items of the NIHSS), and their interactions. We test the normality of the distribution of residuals and the random-effects at each level using the Kolmogorov-Smirnov test. Statistical significance for these effects is based on Type III Wald ${\chi^2}$-tests for the change in deviance ($\alpha=0.05$). These ${\chi^2}$-tests and p-values thus reflect null-hypothesis significance tests controlling for all of the other variables in the model.

\subsection{Trajectory Profile Clustering}
The Trajectory Profile Clustering algorithm \cite{KRI20} is designed to group together patients based on the similarities of their disease trajectories. In essence, it uses graphical tools to generate trajectory profiles for each individual that track their evolution of symptoms across time, then clusters them into communities of similarly behaving individuals that define a recovery subtype. The algorithm proceeds as follows:
\begin{enumerate}
	\item \textit{Model using bipartite networks}: At time point $t$ we construct a bipartite graph modeling connections between $N$ individuals and $V$ disease variables/ symptoms. The connections between the individuals and symptoms are encoded through an adjacency matrix $A_t$ of size $N \times V$. For $M$ time points, we can represent the set of these bipartite graphs as an $N \times V \times M$ by stacking the $A_t$ across time points to generate a tensor $X$ where $X_{ivt}$ gives the value of individual $i$’s disease symptom $v$ at time $t$. 
	\item \textit{Threshold and binarize to obtain trajectory profile}: We threshold each symptom to set values less than a fixed fraction $\kappa$ of the maximum score for the symptom to zero,. For example, if symptom \(\nu\) takes score values in (0,1,2,3,4,5), if $\kappa = 0.5$, we binarize scores such that scores below $5 \times 0.5 = 2.5$ are set to 0, and above are set to 1. We call this thresholdded matrix the trajectory profile matrix, $T^i$ for patient $i$, that contains a representation of how the set of symptoms that a patient is \textit{severely affected} by varies with time. The matrix entries of $T^i$ are calculated as follows:
	\begin{align}
		T^i_{\nu t} &= 0 \text{ if } X_{i \nu t} \leq max(\nu) \cdot \kappa \\
		 &= 1 \text{ if } X_{i \nu t} > max(\nu) \cdot \kappa 
	\end{align}
		
	Since the range of values for each symptom represents the entire scale of severity, this thresholding ensures that patients are only considered `connected' to symptoms that they severely express.

	\item \textit{Create a patient-patient network based on trajectory similarity}: We create a patient-patient network $P$ of all patients. The nodes of this network denote patients, and the strength of a link between patient $i$ and patient $j$ captures the similarity of their trajectory profiles. $P$ has an adjacency matrix given by: 
	
	\begin{eqnarray}
	\label{eq:2}
	P_{ij}=\sum_{v,t} (T^{i}_{vt} \equiv T^{j}_{vt}).
	\end{eqnarray}
	
	In other words, $P_{ij}$  gives the number of matrix entries for which trajectory profile $T^i$ has the same value as $T^j$. This formulation implies that symptoms are \textit{equally weighted}. While the approach is amenable to non-uniform weighting, there is little clinical consensus on the relative importance of different symptoms. Hence, in the interest of not introducing external bias, we choose uniform weighting, adopting an agnostic approach that assumes all symptoms/indicators are equally important. Other applications may require unequal weighting for symptoms and different time points, in which case one may calculate the patient-patient matrix as follows: $P_{ij}=\sum_{v,t} w_{vt}(T^{i}_{vt} \equiv T^{j}_{vt})$ where $w_{vt}$ is the weight of symptom $v$ at time $t$. 
	
	\item \textit{Cluster the network to identify subtypes}: 
	We then perform Louvain community detection \cite{BLO08} to maximize the Newman-Girvan modularity function \cite{GIR02} on the network defined by the adjacency matrix P. Such community detection allows us to identify `communities' of patients, where individuals within a community have a relatively more similar stroke recovery profiles than patients between communities. As is common in network community detection approaches, the number of communities is not set a priori, but rather chosen so that the modularity is maximized. This process allows us to cluster trajectory profiles, and hence patients, into subtypes which have high intra-subtype similarity. The subtypes are denoted by $C^1, C^2, \ldots C^L$ where each $C^l$ is a collection of trajectory profiles of the patients in that subtype, and $L$ is the total number of subtypes.
	
	\item \textit{Construct aggregate profiles to characterize each subtype}:
	We average the trajectory profiles of all patients in each community $C^l$ to obtain the ‘community/subtype profile’ $S^l$. The subtype profile is indicative of the symptom features that describe the subtype. More specifically, it is the normalized average of the trajectory profiles of all the patients in that subtype, i.e., $S^l$ is a $V \times M$ matrix with elements defined by
	\begin{eqnarray}
	\label{eq:3}
	S^l_{vt}=\frac{\sum_{i \in C^l} T^{i}_{vt}}{N_l}
	\end{eqnarray}
	where  $N_l$  is the total number of individuals in subtype $C^l$. 
\end{enumerate}

\subsection{Graph Neural Network}
Graph neural networks (GNNs) \cite{SCA08} are a machine learning approach that captures the relationships represented in graphs through message passing between the nodes of those graphs. GNNs take a graph as input and pass them several layers of nodes, artificial `neurons'. Here we use graph neural networks to identify the timepoints that are most relevant in determining stroke recovery subtypes. Specifically, we train a Graph Neural Network on symptom-symptom graphs generated at each timepoint, and test the accuracy of a GNN in its ability to classify an individual into the correct recovery subtype using data from a single timepoint. A higher accuracy at a given timepoint implies that the recovery subtypes attributed to the patients are strongly correlated with the symptom profiles at that timepoint. 
\\

We generate a symptom-symptom interaction graph $G_t$ at each timepoint $t$ where the nodes represent the disease symptoms. This graph is undirected (i.e., if node $x$ is connected to node $y$, vice-versa is also true, i.e., the adjacency matrix of this graph is symmetric). The graph is generated as follows.
First we generate a symptom-patient binary interaction network for a given timepoint as in step (3) of the previous section. 
We then project it to symptom space to obtain the symptom interaction profile of each patient at a timepoint. The corresponding adjacency matrix (of size $V \times V$) for the graph for patient $i$ is given by
\begin{equation}
    G_{it} = T_{ivt}^T \times T_{ivt}
\end{equation}
Lastly, We repeat this for each individual such that there exist $N$ symptom-symptom interaction graphs at each timepoint.

We then separate the individual cases into a training data set (70\% of total individuals) and a test data set (30\% of total individuals) used to validate our approach.
A convolutional graph neural network is trained on a graph classifying task for each time point, with labels provided by the subtypes/communities of that individual. The stratification of individuals into their recovery subtypes at each timepoint is then measured by testing the accuracy of the GNN on the test data for each timepoint.

The graph neural network takes as input symptom-symptom networks where we consider the 15 NIHSS assessment items as the nodes. The network conists of an input layer, a single hidden layer, and an output layer. The hidden layer comprises of 64 artifical neurons. The input is processed through two graph convolutional layers with ReLU nonlinearities. We then calculate the graph representation by averaging all the neuron representations in the output layer, which contains an equal number of neurons to the number of subtypes. The output is passed through a softmax classifier that yields the probability of the graph belonging to a particular category/subtype. We use cross-entropy loss and an optimizer (ADAM) for adaptive moment estimation.

\section{Results}
\subsection{Key Patterns in Slopes from the Linear Mixed-Effect Models}
\label{results:mem}
First, we present results of the linear mixed-effects-model, a more conventional statistical approach used to model medical data. Modeling each domain of the NIHSS using the mixed-effects model demonstrated several key-patterns in the data. 
First, the level of impairment at the beginning of the trial (i.e., at the intercept) differed by domain. We measure the chi-squared value to test if the differences between domains were statistically significant, and found a main-effect of domain ${\chi^2}(14)>999, p<0.001$, a main-effect of Time, ${\chi^2}(1)=677.48, p<0.001$, and there was a significant Time x Domain interaction, ${\chi^2}(14)=609.42, p<0.001$, showing that impairments in different domains resolved at different rates over time, as shown in Fig. \ref{fig:tPA_data}. On average across time and domains, there was also a statistically significant main-effect of treatment vs placebo, ${\chi^2}(1)=12.79, p<0.001$. There was some evidence for a treatment Group x Domain x Time interaction, ${\chi^2}(14)=20.83, p=0.106$ which, although not statistically significant, is conceptually consistent with prior analyses of this data \cite{FEL2002} showing that the benefits of tPA differ depending on the domain of impairment and the time-post stroke. 

\begin{figure}[htbp!]
  \includegraphics[width=0.45\textwidth]{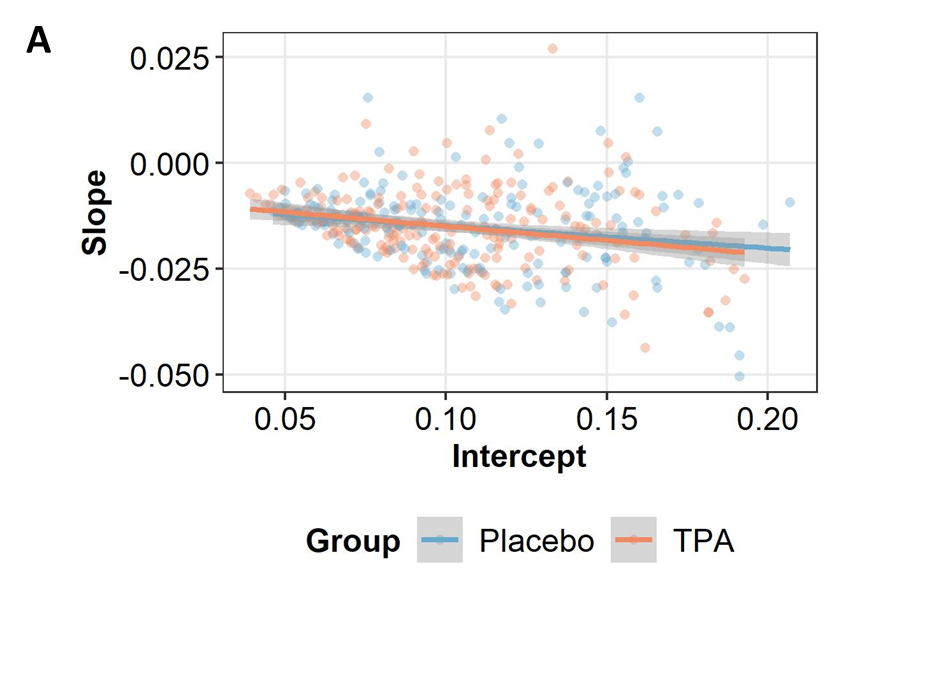}
  \includegraphics[width=0.45\textwidth]{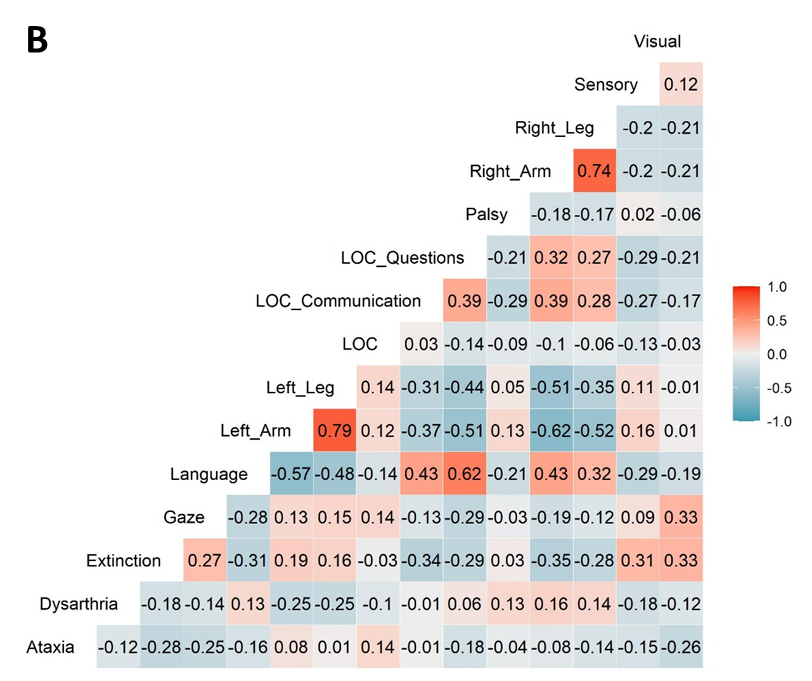}
    \caption{(A) Random-slopes regressed onto the random-intercepts, at the level of individual patients, from the mixed-effects regression model. Regardless of treatment group, there was negative relationship, such that individuals with lower initial impairment exhibited less of a change over time. (B) Correlogram showing the association between slopes in the different domains at the patient:domain level of the model. Correlations are shown as Spearman rank-order correlations. Red boxes indicate positive correlations and blue colors indicate negative correlations.}
  \label{fig:correlations}
\end{figure}

Inspection of the random effects of this model also provides some insights relevant to the current goal of creating behavioral phenotypes. As shown in Fig. \ref{fig:correlations}(A), there was a negative association between slopes and intercepts, on average across the NIHSS assessment items, such that \textit{subjects who had higher levels of initial impairment had more negative slopes}. This negative association is (at least in part) driven by a floor effect in the NIH stroke scale, where zero values indicate an assessment of no impairment. This pattern was generally consistent across the different domains of the NIHSS. 

More importantly, as shown in Fig. \ref{fig:correlations}(B), there were several unique correlations among slopes at the patient:domain level of the model that are of interest: First, as would be expected given the common occurrence of post-stroke hemiplegia (weakening on one side of the body), some of the strongest correlations were for the ipsilateral arm and leg.  The right arm and leg showed a similar time-course of change in impairment ($r=0.74$), as did the left arm and leg ($r=0.79$). Second, there are also patterns in the correlation matrix consistent with lateralization of function as affected by unilateral stroke. For instance, the NIHSS item for extinction was positively associated with left arm/leg deficits (i.e. right hemisphere damage), and negatively associated with right arm/leg deficits (i.e.left hemisphere damage). Gaze was also positively associated with this set of symptoms, albeit to a much smaller degree, possibly reflecting the fact that gaze deviation is typically more pronounced in patients with hemineglect \cite{FRU2006}. Lastly, language was positively associated with the right arm/leg (left hemisphere) and negatively associated with the left arm/leg (right hemisphere). 

The three points outlined above highlight critical insights that can be provided by conventional approaches. However, this linear mixed-effects models is unable to resolve a finer level of detail for the time-evolving interactions that may underlie stroke recovery subtypes. Additionally, limitations in interpreting these data are the non-normal distributions of both the residuals and the random-effects from the model. For instance, a Kolmogorov-Smirnov test of the studentized residuals showed a strong departure from normality, $p<2.2e-10$. Thus, although the marginal effects are informative, this statistical approach is clearly not optimal for understanding how patients' symptoms are changing over time with these ordinal data where strong ceiling/floor effects are present. These limitations are overcome in the network-based TPC method, which is particularly well suited for longitudinal data of any type (e.g. normal, non-normal categorical, ordinal). The agreement between the two approaches, as detailed below, serves to further validate TPC as an effective, flexible and intuitive approach.

\subsection{Stroke Recovery Subtypes identified by TPC}
 Maximizing modularity on the patient-patient trajectory-similarity network gives us three distinct recovery subtypes. It is worth mentioning that the number of subtypes are not predetermined, but are optimally chosen such that the modularity is maximized, i.e., the subtypes are optimally separated.

\begin{figure}[htbp!]
  \includegraphics[width=0.85\textwidth]{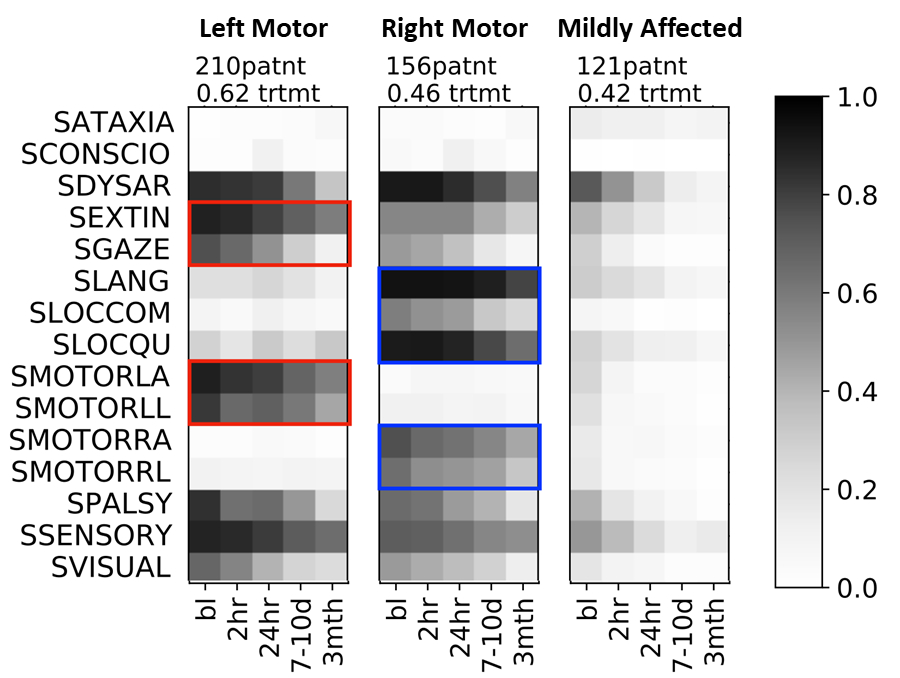}
    \caption{
    	Corresponding profiles of the 3 stroke recovery subtypes. Subtypes identified by the algorithm containing fewer than 10 patients are not shown (1 outlier patient falls under this category). The shade of grey indicates the affected fraction i.e., fraction of patients in the recovery subtype that are severely affected by that symptom at that time. The number of patients in the subtype, and fraction of patients receiving treatment is listed above each panel. The symptom names are listed to the left. The red boxes highlight the unique combination of dominant symptoms of the `left motor' subtype. The blue boxes highlight the unique combination of dominant symptoms of the middle `right motor' subtype. The rightmost `mildly affected' subtype has the mildest symptom profile. The symptoms names on the left are preceded by the letter `S' (indicating `Stroke') to be consistent with the naming convention in the dataset.}
  \label{fig:subtypes}
\end{figure}


Fig.\ref{fig:subtypes} shows the clinical profiles of each subtype. The darkness of the shade of grey for each symptom over time denotes the fraction of patients who had a value above  threshold for that symptom. To reiterate, the thresholding ensures that patients are only considered affected with symptoms for which they display relatively high severity, defined to be above the population median. 
Our analysis identifies 3 distinct stroke trajectory profiles that align with clinically relevant stroke syndromes, characterized both by distinct clusters of symptoms, as well as differing degrees of symptom severity over time. Several key features of the identified subtypes warrant comment. First, our TPC approach identifies a `mildly affected' group that was the least symptomatic of the three subtypes both in terms of baseline severity and 3-month residual symptoms. As a group, this subtype showed a mixture of features that are not clearly lateralizing. In addition, two severely affected subtypes are readily identified that correspond to left and right hemisphere syndromes: We find a `left motor' subtype, showing severely impaired left arm and leg strength together with hemineglect, but with essentially no right-sided motor symptoms (red boxes, Fig.\ref{fig:subtypes}), and a `right motor' subtype, showing severely impaired right arm and right leg strength together with aphasia (blue boxes, Fig.\ref{fig:subtypes}). Additionally, spatial perception scale items are most affected in the `left motor' group (corresponding clinically to a right hemisphere syndrome with hemispatial neglect). Conversely, the language and question-answering items are most affected in the `right motor' group (corresponding to a left hemisphere syndrome with primarily expressive aphasia). These findings are in alignment with prior factor analysis on the clinimetric properties of the NIHSS \cite{LYD1999} and Principal Component Analysis (PCA) to define common behavioral clusters \cite{CORB2015}, as well as results from our mixed-effects model in section \ref{results:mem}. The fact that our results capture clinically relevant subtypes and corroborate these prior findings supports the content validity of this analytic approach. Additionally, our clustering approaches reveal subtype structure at a finer scale (both in terms of symptoms as well as longitudinal symptom evolution) than can be achieved with PCA, and results that are clinically consistent.

It is notable that all three identified trajectory subtypes included both tPA- and placebo-treated patients, suggesting that treatment effects were less defining characteristics of patient recovery profiles than were initial severity and stroke laterality. TPC also provides interesting insights into patterns of symptom prevalance over time across subtypes. Spatial perception deficits (hemineglect) are present in both the left-motor and right-motor subtypes, but tend to be milder and have better resolution in left hemisphere strokes. This observation reinforces the importance of targeted screening during rehabilitation for hemineglect symptoms in both left- and right-hemisphere stroke, since persistent milder symptoms that could be amenable to treatment might otherwise be overlooked. Visual deficits are also present in both left- and right-hemisphere strokes as would be expected, but contrary to conventional understanding that visual deficits resolve less well than hemineglect, the overall prevalence of persistent visual symptoms at 3 months is lower than for hemineglect.  





\subsubsection{Effects of tPA Treatment and Time}

\begin{figure}[htbp!]
  \includegraphics[width=0.45\textwidth]{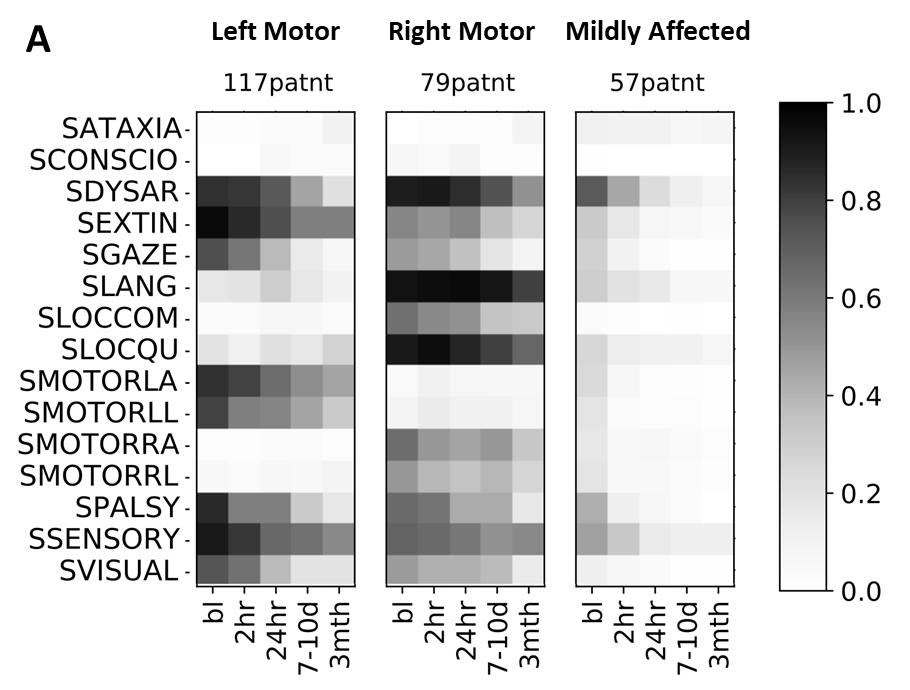}
  \includegraphics[width=0.45\textwidth]{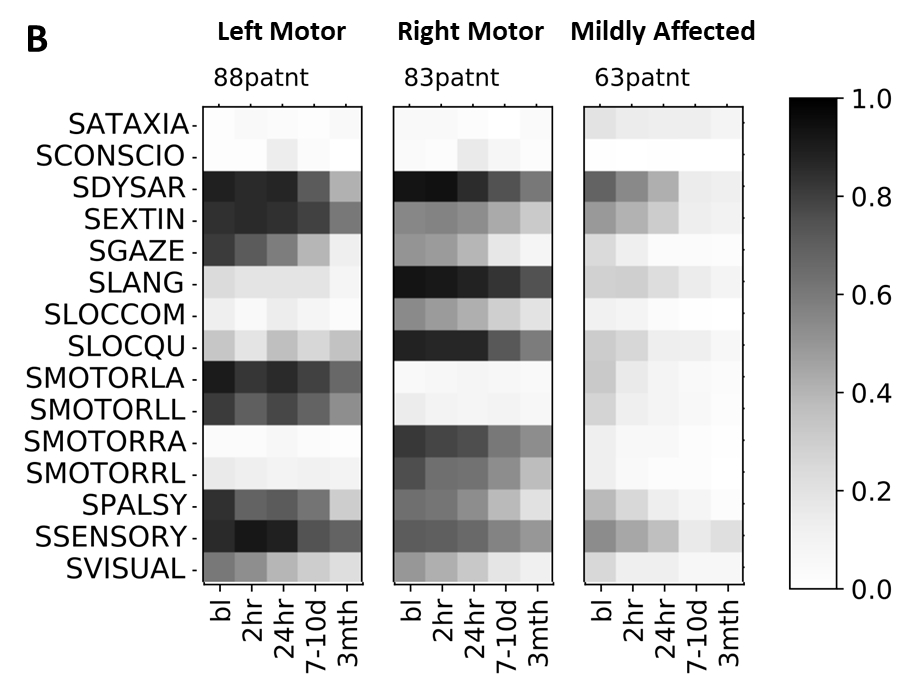}
    \caption{Trajectory profiles (same as in the above figure) were applied independently on patients that (A) received tissue Plasminogen Activator (tPA) treatment within 3-hrs of stroke onset compared to (B) patients who received placebo. Subtypes identified by the algorithm containing fewer than 10 patients are not shown (1 outlier patient falls under this category). The shade of grey indicates the affected fraction.}
  \label{fig:TP_effects}
\end{figure}

\begin{figure}[htbp!]
  \includegraphics[width=0.5\textwidth]{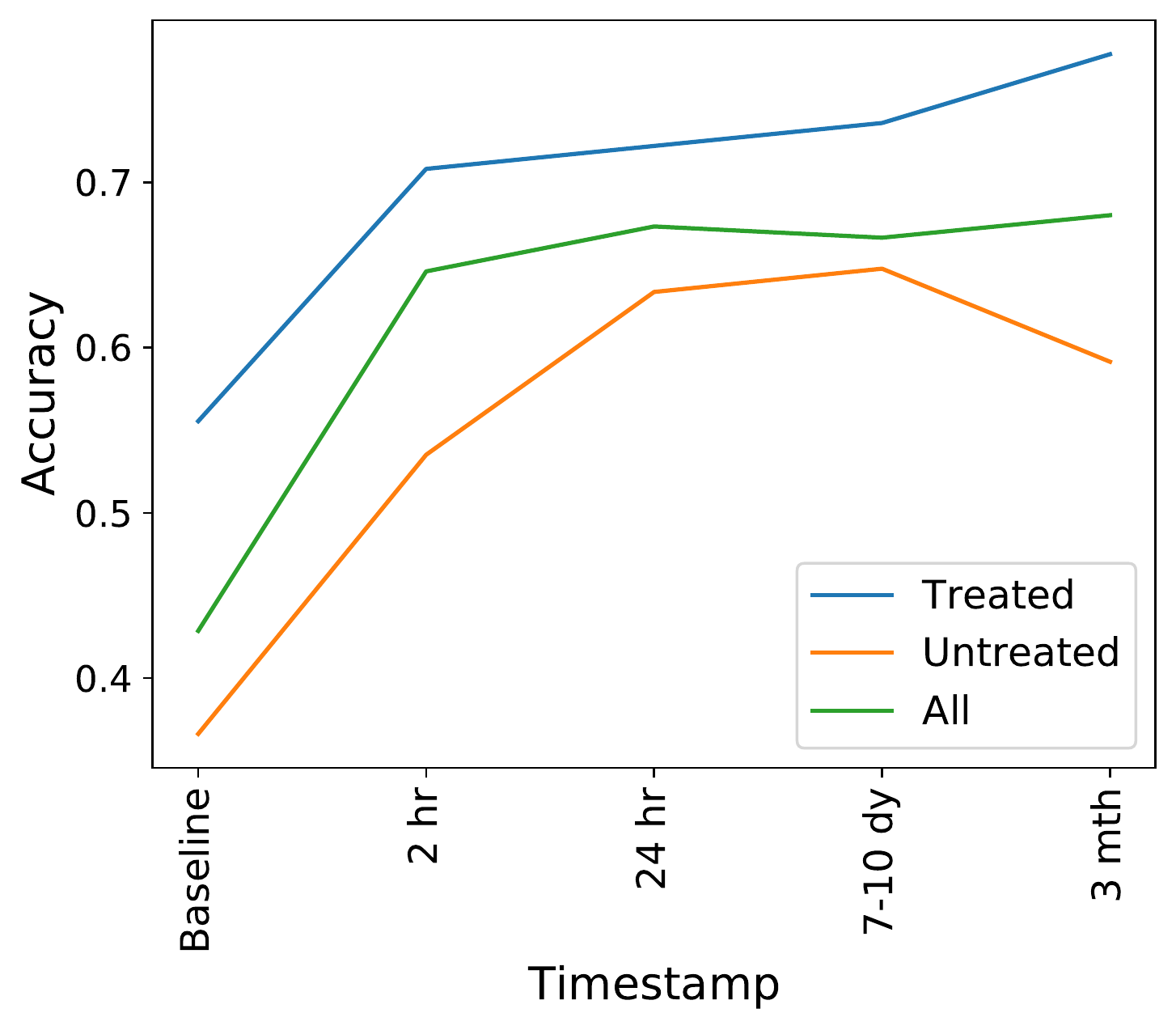}
    \caption{Test accuracy denoting predictive power of a graph neural network as a function of timepoint. Accuracy plotted separately for patients that received tPA treatment, placebo patients and all patients (tPA treated+placebo). 70\% data used for training the neural network, 30\% for testing. Number of training epochs = 100. We use a 2 layer graph convolutional neural network with 16 hidden units and relu nonlinearity at both layers.
    }
  \label{fig:accuracy}
\end{figure}

A natural extension of our TPC subtyping is to study the timecourse of stratification into these subtypes. One might wonder whether subtype (and consequently the expected recovery profile of a patient) is largely driven by baseline symptom severity or by early stroke treatments. Machine learning is particularly well-posed to answer such questions. Since we operate in the graph domain, we use graph neural networks. We first extract trajectory profiles independently on patients who had received tPA within 3 hours of stroke onset compared to patients who had received a placebo. In Fig. \ref{fig:TP_effects} we see that the identified subtypes retain the same symptom clusters identified in Figure Fig.\ref{fig:subtypes}(B), but that overall symptom severity is lower in the tPA-treated population Fig. \ref{fig:TP_effects}(A), particularly in symptoms that are dominant identifiers of the group. For instance, in the left-motor group, assessment items for gaze, left arm and leg strength, and sensation showed higher recovery in the tPA-treated group (Fig. \ref{fig:TP_effects}(A)) versus placebo group (Fig. \ref{fig:TP_effects}(B)). Similarly, in the right-motor group, assessment items for right arm and leg strength, and command-following (SLOCCOM) showed higher recovery in the tPA-treated group (Fig. \ref{fig:TP_effects}(A)) compared to the placebo group (Fig. \ref{fig:TP_effects}(B)). As expected, the effect of tPA is less obvious in the minimally impaired group.
We explore in Fig. \ref{fig:accuracy} the accuracy of a neural network in predicting the subtype of an individual given data at a discrete timepoint.  We generated a symptom-interaction network for each individual at each timepoint, and trained a convolutional-GNN to learn properties of interaction with neighbors.  The convolutions are used for averaging over the neighborhood. If the learned properties for that timepoint are separated according to subtypes, then data at that timepoint is considered a good predictor of the subtype. In the training stage, we assume that the subtype is known to the neural network, which attempts to learn correlations between symptom-interaction patterns and the subtype. We then test to identify if the features learned by the neural network are consistent with the actual subtypes of the test patients. 

Fig. \ref{fig:accuracy} shows that there is a difference in predictive accuracy at baseline for the tPA vs. Placebo groups. The baseline timepoint is a more accurate predictor of subtype for patients who received tPA. This finding may seem unexpected if one posits that tPA `rescues' an otherwise poor prognosis with severe baseline symptoms predicting poor outcomes. However the predictive accuracy for the tPA group rapidly increases during the first 2 hours, and given the expected timecourse for therapeutic effects of tPA, this finding provides additional validation of our approach. For the Placebo group, predictive accuracy grows at comparable rates (comparable slopes) up to the 24 hour mark, showing peak predictive accuracy at the 7-10 day mark, with standard error on the order $10^{-2}$. The tPA group showed a further uptick in predictive accuracy by the 3 month mark, suggesting that treatment continues to exert an effect on recovery subtype stratification even in the later stages of post-stroke. One might speculate that this is the result of tPA treatment salvaging a greater `reserve' of neural tissue for later rehabilitation therapies to act upon. Our report on the rapid increase in predictive accuracy from 2-24 hours post-stroke furthermore aligns with recent work by Heitsch et al. \cite{HEI2021}who reported on the early change in NIHSS scores between 6 – 24 hours as a dynamic phenotype associated with long-term outcomes. 



\section{Conclusion and Future Work}
In this work we introduce a network-based, data-driven method for stroke recovery analysis. First, we analyze the NINDS tPA stroke dataset using conventional quantitative medicine methods including a linear mixed-effects regression model, examining the effects of time, group (tPA vs. Placebo), neurological domain, and their interactions. Further, to identify stroke recovery subtypes and examine their characteristics at a finer resolution, we use the Trajectory Profile Clustering method which accounts not only for symptom severity at different timepoints, but also symptom interactions and their temporal evolution. Of note, although the analytical approach is clinically agnostic, we identify subtypes that are clinically relevant. In particular, we identify a mildly-affected recovery subtype comprising a larger proportion of patients who received tPA. Additionally, we observed that the two other recovery subtypes stratify as left- versus right-sided hemiplegia.  Additionally, we identified that left motor deficits are strongly correlated with deficits in gaze and extinction, whereas right motor deficits correlated with deficits in language. These results again are biologically relevant, and are further validated by convergent findings in the mixed-effects regression model. Lastly, we use graph neural networks to study how much of the stratification into subtypes is identifiable at different time points, and found that stroke recovery trajectories were largely defined within the first 24 hours, consistent with the expected pharmacodynamics of tPA treatment delivered in the first 3 hours after stroke.

This paper is the first work introducing network trajectory approaches for stroke recovery phenotyping, and is aimed at enhancing the translation of such novel computational approaches for practical clinical application. This work presents a data-driven method that is widely applicable to heterogenous neurological disorders such as stroke, and bridges the fields of predictive medicine and network informatics. Because our approach is uniquely adapted to accommodate input variables on multiple scales, future applications could include the integration of other types of data that may contribute to the heterogeneity of recovery, such as data on patient genotypes. 

\begin{backmatter}
\section*{Availability of Data}
  The NINDS dataset was first released in \cite{data} and is publicly available. The code is available on \textit{github/chimeraki/Stroke-Analysis}
  
\section*{Competing interests}
  The authors declare that they have no competing interests.

\section*{Author's contributions}
    SK developed and implemented the TPC algorithm. KL implemented and studied the mixed-effects model on the data. RB conceived, facilitated and guided the project. SK and KL analyzed the computational results and KL and RB interpreted the results in a medical context. All authors were involved in writing the manuscript.

\section*{Acknowledgements}
  We would like to acknowledge Michelle Girvan, Adam de Havenon, Michael M. Binkley, Laura Heitsch, and Alen Delic for helpful conversations and perspectives. We would also like to thank Steven C. Cramer and Arne G. Lindgren for their diligent critical review and comments on the manuscript.


\bibliographystyle{bmc-mathphys} 

\typeout{}
\bibliography{bmc_article}      %

\end{backmatter}
\end{document}